\documentclass[showpacs,preprintnumbers,amsmath,amssymb,prb,twocolumn,footinbib]{revtex4}
\usepackage{graphicx}
\usepackage{bm}
\usepackage{subfigure}
\newcommand{\bo}[1]{\mathbf{#1}}
\newcommand{\bos}[1]{\boldsymbol{#1}}

\newcommand{\ex}[1]{\mathrm{e}^{#1}}
\newcommand{\bom}[1]{\bos{\mathcal{#1}}}
\newcommand{\im}{\mathrm{i}}
\newcommand{\vep}{\varepsilon}

\newcommand{\fv}[1]{\left\langle #1 \right\rangle}
\newcommand{\bra}[1]{\left\langle#1\right|}
\newcommand{\ket}[1]{\left|#1\right\rangle}

\newcommand{\leng}[1]{\left\vert#1\right\vert}

\newcommand{\uu}{\uparrow\uparrow}
\newcommand{\dd}{\downarrow\downarrow}
\newcommand{\ud}{\uparrow\downarrow}
\newcommand{\du}{\downarrow\uparrow}
\newcommand{\bsq}{\bos{q}}
\newcommand{\bsk}{\bos{k}}
\newcommand{\bsp}{\bos{p}}
\newcommand{\de}{\mathrm{d}}

\newcommand{\newatop}[2]{\genfrac{}{}{0pt}{}{#1}{#2}}

\DeclareMathOperator{\tr}{tr}
\DeclareMathOperator{\Tr}{Tr}

\begin{document}
  
\title{Derivation of the Ginzburg-Landau equations
       of a ferromagnetic $p$-wave superconductor}

\author{E. K. Dahl$^{1}$ and A. Sudb{\o}$^{1,2}$}

\affiliation{
  $^{1}$ Department of Physics, Norwegian University 
  of Science and Technology, N-7491 Trondheim, Norway\\
  $^{2}$ Centre for Advanced Study at the Norwegian Academy 
  of Science and Letters, 
  Drammensveien 78, N-0271 Oslo, Norway}

\date{\today}

\begin{abstract}
   We  derive a Ginzburg-Landau free energy for a 
   $p$-wave ferromagnetic superconductor.  The
   starting point is a microscopic Hamiltonian including
   a spin generalised BCS term and a Heisenberg exchange 
   term. We find that coexistence of magnetisation and
   superconductivity depends on the sign of the energy-gradient 
   of the DOS at Fermi level. We also compute the tunneling
   contribution to the Ginzburg-Landau free energy, and find
   expressions for the spin-currents and Josephson currents 
   across a tunneling junction separating two ferromagnetic
   $p$-wave superconductors.
\end{abstract}


\maketitle

\section{\label{sec:intro} INTRODUCTION}
In recent years, experimental evidence for materials
that simultaneously feature superconductivity and 
ferromagnetism, has transpired  from a number of independent studies \cite{pfleiderer2001,aoki2001,saxena2000}. This has triggered much research into this class of materials, widely 
known as ferromagnetic superconductors (FMSC). There are still 
many unanswered questions concerning the properties of these systems featuring  coexisting
spontaneously broken symmetries (broken $U(1)$ and $O(3)$
symmetries). Recently, the superconductivity in ZrZn$_2$
has been found to be surface sensitive, and might in fact 
not be a bulk property of the material \cite{yelland2005}.
A common feature for all the ferromagnetic superconductors 
is that they are superconducting only in the ferromagnetic 
phase. That is, when pressure is increased in such a way as 
to destroy the ferromagnetic phase transition, then superconductivity, if present, is also lost.

Ferromagnetic superconductivity is of great interest from a
theoretical point of view. It offers a laboratory for studying condensed matter systems with multiple 
spontaneously broken symmetries, and their interplay.
 From a technological point of view,
it is also hoped that heterostructures with FMSC will give rise 
to novel transport effects involving both the charge and the 
spin of the electron, which may have the potential
for  being exploited in novel 
types of devices. Recently, structures containing FMSC have 
been investigated theoretically \cite{eremin2005}. This 
work, however,  considers tunneling effects between a spin-singlet superonductor in a Fulde-Ferrel-Larkin-Ovchinnikov state
\cite{FF1964,LO1965} 
coexisting with helimagnetic order.  On the other
hand, little is known of the tunneling effects and interplay between ferromagnetism and superconductivity in spin-triplet
superconductors coexisting with {\it uniform} magnetic order. 
It is the purpose of this paper to derive a Ginzburg-Landau 
model for such a system starting from a reasonable microscopic
Hamiltonian, and then apply the results to computing tunneling
effects both in the charge and spin sector, and thus to elucidate
the interplay between the broken $U(1)$ and $O(3)$ symmetries 
form such systems. Previously, the Bogoliubov-de Gennes equations for such systems have been studied in some detail
\cite{powell2003}.

Phenomenological models designed to describe FMSC were proposed
soon after their experimental discovery \cite{machida2001,walker2002}.  These theories can be written
down from  symmetry considerations of the two order parameters in the problem, i.e. the theories make no reference to the underlying microscopic physics. In ferromagnetic superconductors it is
believed that it is the \emph{same} electrons that are responsible for both superconductivity and ferromagnetism. In such a scenario, the Cooper-pairs must have a magnetic moment, which
means that a FMSC with uniform magnetic order must be 
a spin triplet superconductor. Microscopic theories attempting 
to explain the positive interaction between electrons with aligned spins have traditionally evolved around spin mediated interactions,
i.e. magnon exchange \cite{fay1980,karchev2003}.

In Ref.\onlinecite{shopova2005} the phenomenological model of
Refs. \onlinecite{machida2001,walker2002},  was employed 
in order to find stable solutions and the regions were 
these solutions of the model are stable. The phenomenological 
model used in Refs. \onlinecite{machida2001,walker2002} is given 
by
\begin{align}
\label{eq:walkermodel}
f\left(\psi,\bos{M}\right)=f_S(\psi)+f_F(\bos{M})+f_I(\psi,\bos{M})+\frac{1}{2}\bos{H}^2,
\end{align}
were $\psi$ is a three dimensional complex vector describing
superconducting order,
$\bos{B}=(\bos{H}+4\pi\bos{M})=\nabla\times\bos{A}$ is the magnetic
induction, $\bos{H}$ the external magnetic field ,and $\bos{A}$ is the
electromagnetic vector potential.

The term $f_S(\psi)$ describes superconductivity when
$\bos{H}=\bos{M}=0$, and is given by
{\small
\begin{align}
\label{eq:walkerS}
f_S(\psi)=f_{\text{grad}}(\psi)+a_s\left|\psi\right|^2+\frac{b_s}{2}\left|\psi\right|^4+\frac{u_s}{2}\left|\psi^2\right|^2
+\frac{v_s}{2}\sum_{i=1}^3\left|\psi_i\right|^4,
\end{align}
}
with
\begin{align}
\label{eq:walkerSG}
f_{\text{grad}}(\psi)=&K_1\left(D_i\psi_j\right)^*\left(D_i\psi_j\right)+K_2[(D_i\psi_i)^*(D_j\psi_j)\nonumber
\\
&+(D_i\psi_j)^*(D_j\psi_i)]+ K_3(D_i\psi_i)^*(D_i\psi_i),
\end{align}
where summation over the indices $i,j$ is assumed and the symbol
\begin{equation}
D_i=-\im\frac{\partial}{\partial x_i}+2eA_i,
\end{equation}
of covariant differentiation is introduced. For a discussion of 
the superconductivity part of the free energy see
Refs. \onlinecite{walker2002,shopova2005}. The term $f_F(\bos{M})$ in \eqref{eq:walkermodel} describes the ferromagnetic ordering of the material and is given by,
\begin{align}
\label{eq:walkerF}
f_F(\bos{M})=c_f\sum_{j=1}^3\left|\nabla_jM_j\right|^2+a_f(T_f)\bos{M}^2+\frac{b_f}{2}\bos{M}^4.
\end{align}
This is the standard expression for an isotropic ferromagnet.
Furthermore,
$f_I(\psi,\bos{M})$ describes the interaction between the
order parameters for ferromagnetism and superconductivity, 
and is given by
\begin{align}
\label{eq:walkerI}
f_I(\psi,\bos{M})= \im \gamma_o\bos{M}\cdot(\psi\times\psi^*) 
+ \delta \bos{M}^2\left|\psi\right|^2.
\end{align}

In this paper, we will start out with a microscopic 
Hamiltonian that essentially reproduces the Ginzburg-Landau 
model proposed in Ref. \onlinecite{walker2002}. As a 
byproduct of such a derivation from a more microscopic 
model, we obtain analytic expressions for the coefficients 
in \eqref{eq:walkermodel}. The microscopic theory we start 
with is not trying to explain the {\it mechanism} of the non-unitary spin-triplet pairing of the electron gas. We 
simply assume the existence of some spin generalised attractive 
BCS term in the Hamiltonian, in addition to a Heisenberg ferromagnetic exchange term.  
 
As an application of the Ginzburg-Landau equations, we look at 
two FMSC in tunneling contact. We compute the free energy of 
the coupling and further find the Josephson current and the spin current in the direction perpendicular to the plane spanned by 
the two magnetisation directions.
\section{\label{sec:model} MODEL}
We start out with a Hamiltonian that is given by three terms, one
describing free electrons, one BCS term to account for spin-triplet
superconductivity, and finally  a Heisenberg ferromagnetic
exchange term to account for itinerant ferromagnetism. The Hamiltonian of the system is then given by,
{\small
\begin{align}
\label{eq:hamiltonian}
\hat{H}[c,c^\dagger] = &
\sum_{\bos{k},\sigma}\varepsilon_{\bos{k}}c_{\bos{k},\sigma}^\dagger
c_{\bos{k},\sigma} \nonumber \\
&+
\frac{1}{2}\sum_{ \bos{k},\bos{k}^\prime,\bos{q} }
V_{\bos{k},\bos{k}^\prime}
c_{\bos{k}+\bos{q}/2,\alpha}^\dagger c_{-\bos{k}+\bos{q}/2,\beta}^\dagger
c_{-\bos{k}^\prime+\bos{q}/2,\beta}c_{\bos{k}^\prime+\bos{q}/2,\alpha} \nonumber\\
&+
\frac{1}{2}\sum_{ \bos{q} } J \gamma(\bos{q}) \bos{S}_{\bos{q}} \cdot \bos{S}_{-\bos{q}},
\end{align}
}
where $c_{\bos{k},\sigma}$ and $c_{\bos{k},\sigma}^{\dagger}$ annihilates and creates an electron in the state $(\bos{k},\sigma)$
respectively. Here, $\bos{S}_{\bos{k}}$ is the usual spin operator given by,
 $\bos{S}_{\bos{k}}=\sum_{\bos{q},\alpha,\beta}c_{\bos{k}+\bos{q},\alpha}^{\dagger}\boldsymbol{\sigma}_{\alpha\beta}c_{\bos{q},\beta}$ and
 $\boldsymbol{\sigma}$ is the Pauli matrices. $\gamma(\bos{q})=\sum_{\bos{\delta}}\exp{\im\bos{q}\cdot\bos{\delta}}$ is the structure factor of the underlying lattice,
 and $\bos{\delta}$ is a nearest neighbour vector. A summation 
 over repeated Greek indices is implied.

We are interested in calculating the partition function of the system, formally given  by
\begin{equation}
Z=\Tr\left(\ex{-\beta\hat{H}[c,c^\dagger]}\right)=\ex{-\beta F},
\end{equation}
were $F$ is the exact free energy of the system.
Introducing fermion coherent states, $\xi_{\bos{k},\sigma}$, 
and performing a Hubbard-Stratonovich decoupling of the two 
last terms in \eqref{eq:hamiltonian}, we arrive at an 
effective action (in Euclidean time) which reads,
\begin{widetext}
\begin{align}
S_{\text{eff}}=-\frac{1}{2}\int_0^\beta\mathrm{d}\tau
&\sum_{\bos{k},\sigma}
\left[\xi_{{\bos{k}},\sigma}^{\dagger}\left(\partial_{\tau}+\varepsilon_{\bos{k}}\right)\xi_{{\bos{k}},\sigma}
+\xi_{{\bos{k}},\sigma}\left(\partial_\tau
  -\varepsilon_{\bos{k}}\right)\xi_{{\bos{k}},\sigma}^{\dagger}
\right]\nonumber \\
&+\sum_{\newatop{{\bos{k}},\bos{q}}{\alpha,\beta}}
\left[
\xi_{{\bos{k}}+\bos{q}/2,\alpha}^\dagger\left(\bos{\sigma}\cdot\bos{M}_{\bos{q}}\right)_{\alpha\beta}\xi_{{\bos{k}}-\bos{q}/2,\beta}
-\xi_{-{\bos{k}}-\bos{q}/2\beta}\left(\bos{\sigma}\cdot\bos{M}_{-\bos{q}}\right)_{\alpha\beta}\xi_{-{\bos{k}}+\bos{q}/2\alpha}^\dagger
\right]\nonumber \\
&+\sum_{{\bos{k}},\bos{q}} 
\left[
\Delta_{\alpha\beta}^\dagger({\bos{k}},\bos{q})\xi_{{\bos{k}}+\bos{q}/2,\beta}\xi_{-{\bos{k}}+\bos{q}/2,\alpha}
+\Delta_{\alpha\beta}({\bos{k}},\bos{q})
\xi_{-{\bos{k}}+\bos{q}/2,\beta}^\dagger\xi_{{\bos{k}}+\bos{q}/2,\alpha}^\dagger
\right]\nonumber \\
&-\sum_{\bos{q}}\frac{1}{J\gamma(\bos{q})}\bos{M}_{\bos{q}}\cdot\bos{M}_{-\bos{q}}
-\sum_{{\bos{k}},{\bos{k}}^\prime,\bos{q}}\Delta_{\alpha\beta}^\dagger({\bos{k}}^\prime,\bos{q})V^{-1}_{{\bos{k}}^\prime,{\bos{k}}}\Delta_{\beta\alpha}({\bos{k}},\bos{q})
\end{align}
\end{widetext}
Here, $\Delta_{\beta\alpha}({\bos{k}},\bos{q})$ and $\bos{M}_{\bos{q}}$ are auxiliary fields in the functional
integral representation of the partition function, and 
physically represent the spin-triplet pairing fields and 
magnetisation, respectively. 

The above action is Gaussian in the fermionic fields and hence 
the integral over the fermionic fields may be performed exactly. 
We next introduce a Majorana basis $\bos{\phi}^\dagger_{\bos{k}}=\left[\xi_{{\bos{k}},
    \uparrow}^\dagger\,\xi_{{\bos{k}}, \downarrow}^\dagger \,
  \xi_{-{\bos{k}}, \downarrow} \, -\xi_{-{\bos{k}}, \uparrow}\right]$,
 the effective action may now be written,
{\small
\begin{align}
\label{eq:effaction}
S_{\text{eff}}=-\frac{1}{2}\int_0^\beta\de\tau\Big\{
&\sum_{{\bos{k}},\bos{q}}\bos{\phi}_{{\bos{k}}^\prime}^\dagger
\bos{\mathcal{G}}^{-1}\bos{\phi}_{{\bos{k}}}
-\sum_{\bos{q}}\frac{1}{J\gamma(\bos{q})}\bos{M}_{\bos{q}}\bos{M}_{-\bos{q}}\nonumber\\
-&\sum_{{\bos{k}},{\bos{k}}^\prime,\bos{q}}\tr\bos{\Delta}^\dagger({\bos{k}}^\prime,\bos{q})V^{-1}_{{\bos{k}},{\bos{k}}^\prime}\bos{\Delta}({\bos{k}},\bos{q})
\Big\}.
\end{align}
}
Here,
$\text{tr}$ denotes trace over spin indices. Furthermore, $\bos{\mathcal{G}}^{-1}$
is a $4\times4$ matrix given by, $\bos{\mathcal{G}}^{-1}=\bos{\mathcal{G}}_0^{-1}-\bos{\Sigma}$,
where $\bos{\mathcal{G}}_0^{-1}=
\text{diag}(-\im\omega_n+\vep_{\bos{k}},-\im\omega_n+\vep_{\bos{k}},-\im\omega_n-\vep_{-{\bos{k}}},-\im\omega_n-\vep_{-{\bos{k}}})$ describes a free electron gas.
Here,  $\bos{\Sigma}$ describes the interaction
and pairing correlations, and is explicitly 
given by,
\begin{equation}
\bos{\Sigma}=
\left[
\begin{array}{cc}
\bos{\mathcal{M}}&\bos{\mathcal{D}}\\
\bos{\mathcal{D}}^\dagger&\bos{\mathcal{M}}
\end{array}
\right],
\end{equation}
where $\bos{\mathcal{M}}= \bos{M}\cdot\bos{\sigma}$ is a $2\times2$ matrix
order parameter describing magnetisation and
$\bos{\mathcal{D}}=\bos{d}\cdot\bos{\sigma}=-\im\bos{\Delta}\sigma_y$ is a
$2\times2$ matrix order parameter describing triplet superconductivity.

The integral over $\bos{\phi}$ is Gaussian, and hence it may be performed analytically. The integral produces a fermion 
determinant which may be included in the exponent, i.e. in 
the Ginzburg-Landau free energy, which is given by

\begin{align}
\beta F_{GL}=&- \Tr\ln \bos{\mathcal{G}}^{-1}\nonumber \\
&-\frac{1}{2}\int_0^\beta\de\tau
\left[ 
\sum_{\bos{q}}\frac{1}{J\gamma(\bos{q})}\bos{M}_{\bos{q}}\bos{M}_{-\bos{q}}\right.\nonumber\\
&\left.
 +\sum_{{\bos{k}},{\bos{k}}^\prime,\bos{q}}\tr\bos{\Delta}^\dagger({\bos{k}}^\prime,\bos{q})V^{-1}_{{\bos{k}},{\bos{k}}^\prime}\bos{\Delta}({\bos{k}},\bos{q})
\right],
\end{align}
where $\Tr$ implies a trace over all variables. The GL 
free energy is defined by
\begin{equation}
\int\bom{D}[\xi,\xi^\dagger]\ex{S_{\text{eff}}[\xi,\xi^\dagger]}=\ex{-\beta
  F_{GL}}.
\end{equation}
 The trace over
$\ln \bos{\mathcal{G}}^{-1}$ may formally be
rewritten as
$\Tr\ln\bos{\mathcal{G}}^{-1}=\Tr\ln\bos{\mathcal{G}}_0^{-1}+\Tr\ln\left(1-\bos{\mathcal{G}}_0\bos{\Sigma}\right)$,
the first term, describing free theory, is neglected in the following. The second term is
assumed small, and hence we may expand the logarithm to
obtain
{\small
\begin{align} \Tr\ln
\left(1-\bos{\mathcal{G}}_0\bos{\Sigma}\right)\approx-\Tr\left(\bos{\mathcal{G}}_0\bos{\Sigma}+\frac{1}{2}
\bos{\mathcal{G}}_0\bos{\Sigma}\bos{\mathcal{G}}_0\bos{\Sigma}\right.\nonumber
\\\left.+\frac{1}{3}\bos{\mathcal{G}}_0\bos{\Sigma}\bos{\mathcal{G}}_0\bos{\Sigma}\bos{\mathcal{G}}_0\bos{\Sigma}
+\frac{1}{4}\bos{\mathcal{G}}_0\bos{\Sigma}\bos{\mathcal{G}}_0\bos{\Sigma}\bos{\mathcal{G}}_0\bos{\Sigma}\bos{\mathcal{G}}_0\bos{\Sigma}
+\mathcal{O}(\bos{\mathcal{G}}_0\bos{\Sigma})^5\right)\nonumber\\
=-E_1-E_2-E_3-\dots
\end{align}
}
The first term in the expansion, $E_1=\Tr\bos{\mathcal{G}}_0\bos{\Sigma}$, is zero since the 
Pauli matrices are traceless.The terms $E_2$ and $E_3$ are 
second order in the ordering fields $\Delta_{\beta\alpha}({\bos{k}},\bos{q})$ and $\bos{M}_{\bos{q}}$,
and we now proceed top discuss these terms in turn.

\section{\label{sec:secondorder} Second order term}
The second order term in the expansion of the trace is given as
\begin{align}
E_2=&\frac{1}{2}\Tr\bos{\mathcal{G}}_0\bos{\Sigma}\bos{\mathcal{G}}_0\bos{\Sigma}=
\frac{1}{2}\sum_n\bra{n}\bos{\mathcal{G}}_0\bos{\Sigma}\bos{\mathcal{G}}_0\bos{\Sigma}\ket{n}\nonumber
\\
=&\frac{1}{2}\sum_{\newatop{{\bos{k}}_1,{\bos{k}}_2}{\omega_n}}
\bos{\mathcal{G}}_{0\, \bos{k}_1} 
\bos{\Sigma}_{\bos{k}_1, \bos{k}_2}
\bos{\mathcal{G}}_{0\, \bos{k}_2}
\bos{\Sigma}_{\bos{k}_2,\bos{k}_1}.
\end{align}
The last equality comes from inserting completeness relations and
noting that $\bos{\mathcal{G}}_0$ is local in ${\bos{k}}$-space. We have
introduced the notation $\bos{\mathcal{G}}_{0\,
  \bos{k}_1}=\bra{\bos{k}_1}\bos{\mathcal{G}}_0\ket{\bos{k}_1}$ and
$\bos{\Sigma}_{\bos{k}_1, \bos{k}_2}=\bra{\bos{k}_1}\bos{\Sigma}\ket{\bos{k}_2}$. After
changing variables, $\left.\newatop{\bsk_1}{\bsk_2}\right\}\rightarrow \left\{\newatop{\bsk+\bsq/2}{\bsk-\bsq/2}\right.$, the second order term is
\begin{align}
E_2=\sum_{\newatop{{\bos{k}},\bos{q}}{\omega_n}}\left\{
\left(
g_{0,{\bos{k}}+\bos{q}/2} g_{0,{\bos{k}}-\bos{q}/2}+
c.c
\right)
 \bos{M}_{\bos{q}}\cdot\bos{M}_{-\bos{q}}\right.\nonumber\\
\left.
-\left(g_{0,{\bos{k}}+\bos{q}/2} g^*_{0,{\bos{k}}-\bos{q}/2}
+c.c\right)\left|\bos{d}_{{\bos{k}},\bos{q}}\right|^2
\right\},
\end{align}
where $g^{-1}_{0,{\bos{k}}}=(-\im\omega_n+\varepsilon_{\bos{k}})$. 

The trace of the superconducting order parameter is given 
by
\begin{align}
\tr
\bos{\Delta}^\dagger({\bos{k}},\bos{q})\bos{\Delta}({\bos{k}},\bos{q})=
&\tr\left[\im
  d^\mu_{{\bos{k}},\bos{q}}\bos{\sigma}_\mu\bos{\sigma}_y\right]^\dagger
\left[\im d^\nu_{{\bos{k}},\bos{q}}\bos{\sigma}_\nu\bos{\sigma}_y\right]
\nonumber \\
=&2\bos{d}^*_{{\bos{k}},\bos{q}}\cdot\bos{d}_{{\bos{k}},\bos{q}}.
\end{align}
The complete second order term of the Ginzburg-Landau free energy
is therefore given by
{
\begin{align}
F_2=&
\sum_{\newatop{{\bos{k}},\bos{q}}{\omega_n}}
\Big\{\nonumber \\
&-\Big[
\frac{\delta_{{\bos{k}},0}}{J\gamma(\bos{q})}-
\frac{1}{\beta}
\left(
g_{0,{\bos{k}}+\bos{q}/2} g_{0,{\bos{k}}-\bos{q}/2}+
c.c
\right)\Big]
\bos{M}_{\bos{q}}\bos{M}_{-\bos{q}}\nonumber\\
&-\Big[
\frac{1}{\beta}
\big(
g_{0,{\bos{k}}+\bos{q}/2}
g^*_{0,{\bos{k}}-\bos{q}/2}+c.c\big)
+2/V\Big]
\left|\bos{d}_{{\bos{k}},\bos{q}}\right|^2
\Big\}
\end{align}
}
Now $\bos{q}$ is assumed to be small, and hence we may expand
$\varepsilon_{{\bos{k}}+\bos{q}/2}\approx\varepsilon_{\bos{k}}+\bos{q}/2\cdot
\frac{\partial\varepsilon_{\bos{k}}}{\partial{\bos{k}}}+\mathcal{O}\bos{q}^2=\varepsilon_{\bos{k}}+\bos{q}/2\cdot\bos{v}_F$,
the inverse electron propagator is to first order in $\bos{q}$ given by $g^{-1}_{0,{\bos{k}}+\bos{q}/2}\approx
\left(-\im\omega_n+\varepsilon_{\bos{k}}+\bos{q}/2\cdot\bos{v}_F\right)$.
Hence, we find for the part containing magnetisation
\begin{align}
&g_{0,{\bos{k}}+\bos{q}/2}g_{0,{\bos{k}}-\bos{q}/2}+g^*_{{\bos{k}}+\bos{q}/2}g^*_{{\bos{k}}-\bos{q}/2}\nonumber
\\
=&2\Re\left(\frac{1}{\left(-\im\omega_n+\varepsilon_{\bos{k}}\right)^2}
-\frac{\left(\bos{q}/2\bos{v}_F\right)^2}{\left(-\im\omega_n
    +\varepsilon_{\bos{k}}\right)^4}\right)
\end{align}
when keeping terms to second order in $\bos{q}$. In addition, 
we expand the structure factor $\gamma(\bos{q})\approx 6-2\bos{q}^2$, and assume for simplicity a cubic lattice in 
three dimensions. Now, we substitute $\sum_{\bos{k}}\rightarrow N_0\int_{-\epsilon_c}^{\epsilon_c}\de\xi\int\frac{\de\Omega}{4\pi}$, where $N_0$ is the density of states at Fermi level and $\epsilon_c$ is some cutoff. The homogeneous part of the magnetisation is given by
\begin{align}
F_{2,m}^c=\frac{\beta}{2}\sum_{\bos{q}}\left(4 N_0\tanh\left(\frac{\epsilon_m\beta}{2}\right)+\frac{1}{6J}\right)
\bos{M}_{\bos{q}}\bos{M}_{-\bos{q}}.
\end{align}
The part containing derivatives is given by
\begin{widetext}
\begin{align}
F_{2,m}^d=-\frac{\beta}{2}\sum_{\bos{q}}
\left\{\frac{
N_0(v_F\beta)^2\epsilon_m^3 }{
72}
\tanh\left(
\frac{
\epsilon_m\beta }{
2}
\right)
\left[
1-\tanh^2(\frac{\epsilon_m\beta}{2})^2
\right]
+\frac{1}{36 J}
\right\}\bos{q}^2 \bos{M}_{\bos{q}}\bos{M}_{-\bos{q}}
\end{align}
\end{widetext}

Similarly, we find for the superconducting order parameter, when
introducing $d_\mu({\bos{k}},\bos{q})=\mathcal{A}_{\mu
  i}(\bos{q})\hat{k}_i$\footnote{Here we are assuming that
  the superconducting order parameter is in a $p$-wave state} and the
indices $\mu$ and $i$ run from 1 to 3, that the coefficient in front of
the term involving the superconducting order parameter is given by
\begin{align}
&\sum_{{\bos{k}},\omega_n}2\Re g_{0,{\bos{k}}+\bos{q}/2}g_{0,{\bos{k}}-\bos{q}/2}^*
\hat{k}_i\hat{k}_j\nonumber\\
\approx&\sum_{\omega_n}\int\de\xi\int\frac{\de\Omega}{4\pi}\frac{2N_0\hat{k}_i\hat{k}_j}{\omega_n^2+\xi^2}
\left[
1-\frac{(v_F/2)^2
  q_lq_m\hat{k}_l\hat{k}_m}{\omega_n^2+\xi^2}\right.\nonumber 
\\ & \qquad \qquad\left.+
\frac{\bos{q}/2\cdot\bos{v}_F}{\im\omega_n+\xi}
-\frac{\bos{q}/2\cdot\bos{v}_F}{-\im\omega_n+\xi}
\right].
\end{align}
Here, we have used $\bos{v}_F=v_F\hat{{\bos{k}}}$. The terms 
linear in $\bos{q}$ will integrate to zero when integrated 
over the angles $\Omega$.

We start with the constant term( independent of $\bos{q}$)
\begin{align}
\alpha_c=&\sum_{\omega_n}\int\de\xi\int\frac{\de\Omega}{4\pi}\frac{2N_0\hat{k}_i\hat{k}_j}{\omega_n^2+\xi^2}\nonumber
\\
=&\frac{2\beta N_0\delta_{i,j}}{3}\sum_{n\geq0}^\vep\frac{1}{n+1/2}
\end{align}
where $\vep$ is an energy cut-off. In total the constant second order term is given by
\begin{equation}
F_{2,S}^c=\frac{1}{2}\sum_{\bos{q}}\left(
\frac{2}{3V}-\frac{4N_0}{3}\sum_{n\geq0}^\vep\frac{1}{n+1/2}
\right)\tr\left(\bom{A}\bom{A}^\dagger\right).
\end{equation}
After using the result for the critical temperature \cite{mineev1999},
the term may be written on the form
\begin{equation}
F_{2,S}^c=\frac{1}{2}\sum_{\bos{q}}
\frac{4N_0}{3}\frac{T-T_c}{T_c}
\tr\left(\bom{A}\bom{A}^\dagger\right).
\end{equation}

The coefficient for the second order term containing derivatives 
is given by
\begin{align}
\alpha_d=&-2N_0(v_F/2)^2q_iq_j\sum_{\omega_n}\int\de\xi\int\frac{\de\Omega}{4\pi}
\frac{\hat{k}_i\hat{k}_j\hat{k}_l\hat{k}_m}{\left(\omega_n^2+\xi^2\right)}
\nonumber \\
=&-2N_0(v_F/2)^2q_iq_j1/15(\delta_{ij}\delta_{lm}+\delta_{il}\delta_{jm}+\delta_{im}\delta_{jl})\nonumber\\
&\qquad\qquad
\sum_{\omega_n}\int\de\xi\frac{1}{\left(\omega_n^2+\xi^2\right)}\nonumber
\\
=&-\pi/2 2N_0(v_F/2)^2q_iq_j1/15(\delta_{ij}\delta_{lm}+\delta_{il}\delta_{jm}+\delta_{im}\delta_{jl})\nonumber
\\
&\qquad\qquad\sum_{\omega_n}\frac{\text{sgn}\omega_n}{\omega_n^3}\nonumber
\\
=&-\frac{7q_iq_jv_F^2N_0\beta^3\zeta(3)}{
  240\pi^2}(\delta_{ij}\delta_{lm}+\delta_{il}\delta_{jm}+\delta_{im}\delta_{jl}),
\end{align}
so that the total second order term containing derivatives is given
by
\begin{align}
F_{2,S}^d=&-\frac{7v_F^2N_0\beta^3\zeta(3)}{120\pi^2}\nonumber\\
&\left[
q^2\tr\bom{A}\bom{A}^\dagger+
\left(q_i\mathcal{A}_{\mu  i}\right)\left(q_j\mathcal{A}^\dagger_{j\mu}\right)
+\left(q_j\mathcal{A}_{\mu  i}\right)\left(q_i\mathcal{A}_{j\mu}^\dagger\right)
\right].
\end{align}
Here we have assumed an isotropic quadratic dispersion relation, hence
the coefficients in \eqref{eq:walkerSG} are all the same to this
approximation. If one introduces an effective inverse mass tensor
$(1/m^*)_{i,j}$ one can use the result above by only replacing $ q_i
\rightarrow q_jv_{F i,j}$, where $v_{F i,j}=
1/2[(1/m^*)_{i,j}+(1/m^*)_{j,i}]$, and remove $v_F$. In this way space
is no longer isotropic.
\section{\label{sec:thirdorder} Third order term}
In this section we will find a coupling between the
order parameters for magnetisation and superconductivity. The third order
term will, however, only be nonzero when the system is in a non unitary
phase, i.e. when $\bos{d}^*\times\bos{d}\neq0$. In a ferromagnetic
superconductor where the \emph{same} electrons are responsible for
superconductivity and ferromagnetism, the system, however, has to be in a non-unitary state. If the system is in a unitary state the superconducting order parameter is invariant under the time inversion operator, and hence there can not be any magnetism 
associated with it.

The third order term in the expansion of the fermion determinant 
is given by
\begin{align}
E_3=\frac{1}{3}\sum_{{\bos{k}}_1,{\bos{k}}_2,{\bos{k}}_3}
&\bos{\mathcal{G}}_{0\,\bos{k}_1}
\bos{\Sigma}_{\bos{k}_1,\bos{k}_2}
\bos{\mathcal{G}}_{0\,\bos{k}_2}
\bos{\Sigma}_{\bos{k}_2,\bos{k}_3}
\bos{\mathcal{G}}_{0\,\bos{k}_3}
\bos{\Sigma}_{\bos{k}_3,\bos{k}_1}.
\end{align}
Via a change of variables, 
$ \left.\newatop{\bsk_1}{\bsk_2}\right\}\rightarrow\left\{\newatop{\bsk+\bsq_1/2}{\bsk-\bsq_1/2}\right.$, 
$
\left.\newatop{\bsk_2}{\bsk_3}\right\}\rightarrow\left\{\newatop{\bsk+\bsq_2/2}{\bsk-\bsq_2/2}\right.$
and 
$ \left.\newatop{\bsk_3}{\bsk_1}\right\}\rightarrow\left\{\newatop{\bsk+\bsq_3/2}{\bsk-\bsq_3/2}\right.$  along with the constraint, $\bsq_1+\bsq_2+\bsq_3=0$, a multipication of 
the matrices yields the following expression  for the
third order term
\begin{align}
\label{eq:3.coeff}
E_3=4\sideset{}{^\prime}\sum_{\bos{k},\{ \bsq_{i} \} }&
\left(
g_{0,{\bos{k}}}g_{0,{\bos{k}}}^*g_{0,{\bos{k}}}
+c.c
\right) \nonumber \\
&
\im\bos{d}^*_{{\bos{k}},-\bos{q}_1}\times\bos{d}_{{\bos{k}},\bos{q}_3}\cdot\bos{M}_{\bos{q}_2},
\end{align}
were the combination $\bos{m}=\im \bos{d}\times\bos{d}^*$ is
interpreted as the average magnetisation due to the Cooper-pairs
\cite{mineev1999,leggett1975}.
The prime on the sum in Equation \eqref{eq:3.coeff} denotes sumation
over configurations with the restriction $\delta(\bos{q}_1+\bos{q}_2+\bos{q}_3)$.

In the coefficient we have neglected the $\bos{q}$ dependence, 
since expanding in powers of $\bos{q}$ amounts to finding derivatives. The coefficient is given by, when introducing
$\bos{d}_\mu=\mathcal{A}_{\mu i}\hat{k}_i$,
\begin{align}
\tilde{\alpha}_3/3!=
&4\Re\sum_{\omega_n,{\bos{k}}}
\frac{1}{\omega_n^2+\varepsilon_{\bos{k}}^2}
\frac{\hat{k}_i\hat{k}_j}{-\im\omega_n+\varepsilon_{\bos{k}}}\nonumber
\\
\approx&4 N^\prime_0
\sum_{\omega_n}\int_{-\vep}^\vep\de\xi\int\frac{\de\Omega}{4\pi}
\frac{\xi}{\omega_n^2+\xi^2}
\frac{\hat{k}_i\hat{k}_j}{-\im\omega_n+\xi}\nonumber\\
=&N^\prime_0\beta\delta_{ij}/3\sum_{n\geq0}\frac{1}{n+1/2}.
\end{align}
The final summation in the above expression is proportional to the logarithm of the cut-off frequency. The total third order term 
is thus given by
\begin{equation}
F_3=-\sideset{}{^\prime}\sum_{\bos{q}_1,\bos{q}_2,\bos{q}_3}\frac{\alpha_3}{3!}\im
\varepsilon_{\mu\nu\lambda}\mathcal{A}_{\mu
  i}\mathcal{A}^*_{\nu i}M_\lambda.
\label{third_order_term}
\end{equation}
were $\alpha_3=\tilde{\alpha}_3/\beta$.
As the order parameters $\bos{M}$ and $\bos{d}$ do not couple 
to second
order in the GL free energy, the coupling in the third order term
is expected to be of crucial importance as to whether the two order
parameters will coexist in the system or not. Coexistence is favoured by the system if there is an energy
gain by having $\bos{M}$ and $\bos{d}$ finite simultaneously.
Hence, coexistence of magnetism and superconductivity
depends on the sign of $\alpha_3$, which again is given by the
gradient of the DOS at Fermi level. In the system considered 
here, there will be a ferromagnetic coupling between magnetism 
and the spin magnetism of the Cooper-pairs if $N_0^\prime>0$, 
i.e the gradient of the DOS should be positive at the Fermi 
level for the spin-up sheet if coexistence of FM and SC is preferred by the material. 

Comparing  Eq. \eqref{third_order_term} with the model Eq. \eqref{eq:walkerI}, we find
the value of the coefficient $\gamma_0$ when we assume that $\psi$
is in a $p$-wave state. The sign dependence of $N_o^\prime$ is however
more general, since a triplet state must necessarily be an odd
function of $\bos{k}$. From \eqref{eq:3.coeff}, we observe that 
in an expansion of the DOS around Fermi level, 
$N(\xi)\approx N_0+\xi
N_0^\prime+\xi^2/2 N_0^{\prime\prime}+\dots$, only the odd terms of
the expansion give a contribution to the coefficient. Hence, to 
lowest order $\gamma_0 \backsim N_0^\prime$. We next determine 
the fourth order terms in the order-paarameter expansion of the
GL free energy. 

\section{\label{sec:fourthorder} Fourth order term}
In this section, we will find three different types terms, namely
 a term involving only magnetisation, one onoly involving
 superconductivity, and finally one term involving a
coupling between the magnetisation $\bos{M}$ and 
the superconducting orderparameter $\bos{d}$. In total, we will find five different independent terms for a $p$-wave
ferromagnetic superconductor. The fourth order term of the expansion of the fermion determinant is given by
\begin{align}
E_4=\frac{1}{4}\sum_{\newatop{{\bos{k}}_1,{\bos{k}}_2}{{\bos{k}}_3,{\bos{k}}_4}}
&\bos{\mathcal{G}}_{0\, \bos{k}_1}
\bos{\Sigma}_{\bos{k}_1,\bos{k}_2}
\bos{\mathcal{G}}_{0\, \bos{k}_2}
\bos{\Sigma}_{\bos{k}_2,\bos{k}_3}\nonumber\\
&\bos{\mathcal{G}}_{0\,\bos{k}_3}
\bos{\Sigma}_{\bos{k}_3,\bos{k}_4}
\bos{\mathcal{G}}_{0\,\bos{k}_4}
\bos{\Sigma}_{\bos{k}_4,\bos{k}_1}.
\end{align}
As before, we change variables and multiply out the
matrices and take the trace. In addition, we 
make the approximation of 
neglecting the dependence on
$\bos{q}_i$ in $g_{0,{\bos{k}}_i}$.

\subsection{Terms containing only magnetic order parameter}
The term involving the magnetisation only is given by
\begin{align}
\beta F_{4,m}=\sideset{}{^\prime}\sum_{\{\bos{q}_i\}}&\left(\bos{M}_{\bos{q}_1}\cdot\bos{M}_{-\bos{q}_2}\right)\left(\bos{M}_{\bos{q}_3}\cdot\bos{M}_{-\bos{q}_4}\right)
\nonumber\\&\left(\Re\sum_{{\bos{k}},\omega_n} g_{0,{\bos{k}}}g_{0,{\bos{k}}}g_{0,{\bos{k}}}g_{0,{\bos{k}}}\right).
\end{align}
Here, the coefficient of the $\bos{M}_{\bos{q}}$-factors is 
given by the trace over the electron propagators,
\begin{align}
\tilde{\alpha}_{4m}=&\Re\sum_{{\bos{k}},\omega_n}
\left(g_{0,{\bos{k}}}\right)^4=N_0\Re\sum_{\omega_n}\int_0^{\epsilon_{m}}\de\xi\left(\frac{1}{-\im\omega_n+\xi}\right)^4\nonumber
\\
=&\frac{\beta^3 2}{4!}\tanh\left(\frac{\epsilon_{m}\beta}{2}\right)\left[1-\tanh^2\left(\frac{\epsilon_{m}\beta}{2}\right)\right]
\end{align}
Thus, the fourth order term involving magnetisation only, is
given by
\begin{equation}
F_{4,m}=
\sideset{}{^\prime}\sum_{\{\bos{q}_i\}}\frac{\alpha_{4m}}{4!}\left(\bos{M}_{\bos{q}_1}\cdot\bos{M}_{-\bos{q}_2}\right)
\left(\bos{M}_{\bos{q}_3}\cdot\bos{M}_{-\bos{q}_4}\right)
\end{equation}
where $\alpha_{4m}=4!\tilde{\alpha}_{4m}/\beta$.

\subsection{Terms containing only superconducting order parameter}
The term involving the superconducting order parameter only will
contain five different terms. Again neglecting the $\bos{q}$ dependence in $g_{0,{\bos{k}}}$, we arrive at
\begin{align}
\beta F_{4,S}=&\sideset{}{^\prime} \sum_{\{\bos{q}_i\}}\Re
\sum_{{\bos{k}},\omega_n}
\left(
\frac{
\hat{k}_i\hat{k}_j\hat{k}_l\hat{k}_m
}{
(\omega_n^2+\vep_{\bos{k}}^2)^2
} 
\right)\nonumber \\
&\left[
2\mathcal{A}_{\mu i}\mathcal{A}^*_{\mu
  j}\mathcal{A}_{\nu l}\mathcal{A}^*_{\nu m}
-\mathcal{A}_{\mu i}\mathcal{A}_{\mu
  j}\mathcal{A}^*_{\nu l}\mathcal{A}^*_{\nu m}
\right].
\end{align}
Now let the sum over ${\bos{k}}$ go over to an integral over $\xi$ and the
angles $\theta$ and $\phi$. The integral over the angles produces
Kronecker-$\delta$'s in  Latin indices,
\begin{equation}
\int\frac{\de\Omega}{4\pi}\hat{k}_i\hat{k}_j\hat{k}_l\hat{k}_m
=\frac{1}{15}\left(\delta_{i,j}\delta_{l,m}+\delta_{i,l}\delta_{j,m}+\delta_{i,m}\delta_{j,l}\right).
\end{equation}
The coefficient is given by
\begin{equation}
\frac{\alpha_{4S}}{4!}=\frac{N_0}{15\beta}\int\de\xi\sum_{\omega_n}\left(\frac{1}{\omega_n^2+\xi^2}\right)^2=
\frac{7N_0 \zeta(3)}{120\pi^2}\beta^2
\end{equation}
In total, the fourth order term involving superconductivity alone is
given by,
\begin{align}
F_{4,S}=\frac{\alpha_{4S}}{4!}&\sideset{}{^\prime}\sum_{\{\bos{q}_i\}}
\bigg\{
-\left|\tr \bos{\mathcal{A}}\bos{\mathcal{A}}^T\right|^2 
+2\left(\tr
  \bos{\mathcal{A}}\bos{\mathcal{A}}^\dagger\right)^2\nonumber \\
&+2\tr\left(\bos{\mathcal{A}}\bos{\mathcal{A}}^T\right)\left(\bos{\mathcal{A}}\bos{\mathcal{A}}^T\right)^*
+2\tr\left(\bos{\mathcal{A}}\bos{\mathcal{A}}^\dagger\right)^2\nonumber \\ 
&-2\tr
\left(\bos{\mathcal{A}}\bos{\mathcal{A}}^\dagger\right)\left(\bos{\mathcal{A}}\bos{\mathcal{A}}^\dagger\right)^*
\bigg\},
\end{align}
where $\alpha_{4S}=\frac{7\zeta(3)N_0}{5\pi^2}\beta^2$. This 
part of the free energy thus  consists of five independent 
terms \cite{barton1974}.

\subsection{Terms containing  magnetic and superconducting order parameter}
The fourth order coupling term between the magnetisation
and the superconducting order parameter is given by
\begin{widetext}
\begin{align}
\label{eq:4.coupling}
F_{4,Sm}=\sideset{}{^\prime}\sum_{{\bos{k}},\{\bos{q}_i\},\omega_n}\Bigg\{&8g_{0,{\bos{k}}}g_{0,{\bos{k}}}g^*_{0,{\bos{k}}}g^*_{0,{\bos{k}}}
\left[
2\left(\bos{M}_{\bos{q}_1}\bos{d}_{-\bos{q}_2}\right)
\left(\bos{M}_{\bos{q}_3}\bos{d}^*_{-\bos{q}_4}\right)
-\left(\bos{M}_{\bos{q}_1}\bos{M}_{\bos{q}_3}\right)
\left(\bos{d}_{-\bos{q}_2}\bos{d}^*_{-\bos{q}_4}\right)
\right]\nonumber \\
-&16
g_{0,{\bos{k}}}g_{0,{\bos{k}}}g_{0,{\bos{k}}}g^*_{0,{\bos{k}}}\left(\bos{d}_{\bos{q}_3}\bos{d}^*_{-\bos{q}_4}\right)
\left(\bos{M}_{\bos{q}_1}\bos{M}_{-\bos{q}_2}\right)
\Bigg\}\nonumber
\\
=\sideset{}{^\prime}\sum_{\{\bos{q}_i\}}&\frac{7N_0\zeta(3)\beta^3}{3\pi^2}
\left(M_\mu\mathcal{A}_{\mu j}\right)
\left(M_\nu\mathcal{A}^*_{\nu j}\right)
\end{align}
\end{widetext}
where the constant in the free energy is
$\alpha_{4Sm}=72  N_0\zeta(3)\beta^2/\pi^2$.
This coupling term between magnetism and superconductivity 
differs from the fourth order coupling in Eq.
\eqref{eq:walkerI}. In Eq.  the coupling term is proportional 
to $\left|\bos{M}\cdot\psi\right|^2$, whereas we find 
a term proportional to 
$(\bos{M}\cdot\bos{M})\left|\psi\right|^2$. In addition, the
coefficient is positive rather than having an indefinite sign as
commented on in Ref.\onlinecite{shopova2005}. We have, however, assumed that the system is in a $p$-wave state. In the a case of 
a general spin-triplet state, both types of fourth order coupling terms between magnetism and superconductivity may be present, as 
also seen from Eq. \eqref{eq:4.coupling}.

\subsection{Complete fourth order term}  
In total, the fourth order term is thus given by
\begin{widetext}
\begin{align}
F_4=1/4!\sum_{\{\bos{q}_i\}}\Bigg\{&\alpha_{4m}\left(\bos{M}_{\bos{q}_1}\cdot\bos{M}_{-\bos{q}_2}\right)
\left(\bos{M}_{\bos{q}_3}\cdot\bos{M}_{-\bos{q}_4}\right)+\alpha_{4Sm}
\left(M_\mu\mathcal{A}_{\mu j}\right)
\left(M_\nu\mathcal{A}^*_{\nu j}\right)
\nonumber \\
+&\alpha_{4S}\bigg[
-\left|\tr \bos{\mathcal{A}}\bos{\mathcal{A}}^T\right|^2 
+2\left(\tr
  \bos{\mathcal{A}}\bos{\mathcal{A}}^\dagger\right)^2
+2\tr\left(\bos{\mathcal{A}}\bos{\mathcal{A}}^T\right)\left(\bos{\mathcal{A}}\bos{\mathcal{A}}^T\right)^*
+2\tr\left(\bos{\mathcal{A}}\bos{\mathcal{A}}^\dagger\right)^2\nonumber \\ 
&-2\tr
\left(\bos{\mathcal{A}}\bos{\mathcal{A}}^\dagger\right)\left(\bos{\mathcal{A}}\bos{\mathcal{A}}^\dagger\right)^*
\bigg]
\Bigg\}
\end{align}
All in all the Ginzburg-Landau free energy for a ferromagnetic superconductor is given by
\begin{align}
F_{GL}=\int\de^3\bos{r}\Big\{&
\frac{\alpha_S(T)}{2}\tr\bos{\mathcal{A}}\bos{\mathcal{A}^\dagger} +
\frac{\beta_S}{2}\left(D^2\tr\bos{\mathcal{A}}\bos{\mathcal{A}^\dagger} +D_i\mathcal{A}_{\mu i}D_j\mathcal{A}_{j\mu}^\dagger+D_j\mathcal{A}_{\mu i}D_i\mathcal{A}_{j\mu}\right)
\nonumber\\
&+\frac{\alpha_m(T)}{2}\bos{M}\cdot\bos{M}+\frac{\beta_m}{2}\tilde{D}^2\bos{M}\cdot\bos{M}+\frac{\alpha_3}{3!}\im\vep_{\mu\nu\lambda}\mathcal{A}_{\mu i}\mathcal{A}_{\nu i}M_\lambda
\nonumber\\
&+\frac{\alpha_{4S}}{4!}\left[ 2(\tr\bos{\mathcal{A}}\bos{\mathcal{A}^\dagger})
-\left|\tr\bos{\mathcal{A}}\bos{\mathcal{A}^\dagger}\right|
+2\tr (\bos{\mathcal{A}}\bos{\mathcal{A}}^T) (\bos{\mathcal{A}}\bos{\mathcal{A}}^T)^*
+2\tr(\bos{\mathcal{A}}\bos{\mathcal{A}^\dagger})^2
-2\tr(\bos{\mathcal{A}}\bos{\mathcal{A}}^\dagger)(\bos{\mathcal{A}}\bos{\mathcal{A}}^\dagger)^* 
\right]\nonumber\\
&+\frac{\alpha_{4m}}{4!}\left( \bos{M}\cdot\bos{M}\right)^2
+\frac{\alpha_{4ms}}{4!}(M_{\mu}\mathcal{A}_{\mu i})(M_\nu\mathcal{A}_{\nu i}^*)
\Big\},
\end{align}
\end{widetext}
where $D=\nabla+2\im e\bos{A}$, $\tilde{D}=\nabla+\im e\bos{A}$ and $\bos{A}$ is the electromagnetic vector potential.
%
\section{\label{sec:tunneling}Tunneling}
As an application of the Ginzburg-Landau theory derived above, 
we will consider tunneling between two ferromagnetic p-wave superconductors. The Hamiltonian we use is the common choice 
when studying tunneling between systems in equilibrium
\begin{equation}
H  =   H_L + H_R + H_T,
\end{equation}
where $H_{L(R)}$ is given by \eqref{eq:hamiltonian} and 
\begin{equation}
H_T=\sum_{\bos{k},\bos{p}}
\left\{
T_{\bos{k},\bos{p}}^{\alpha,\beta}c_{\bos{k},\alpha}^\dagger d_{\bos{p},\beta}
+T_{\bos{k},\bos{p}}^{* \alpha,\beta}d_{\bos{p},\beta}^\dagger c_{\bos{k},\alpha}
\right\}.
\end{equation}

For a general Hamiltonian, it is straightforward to show that 
the part of the Ginzburg-Landau free energy that contains the tunneling elements, is given by
\begin{equation}
\label{eq:freejunk}
F_J=-\frac{1}{\beta}\Tr \ln\left(1-\bos{\mathcal{G}}_{\rm L}\bos{\mathcal{T}}\bos{\mathcal{G}}_{\rm R}\bos{\mathcal{T}}\right),
\end{equation}
where $\bos{\mathcal{G}}_{\rm L(R)}$ are the Green`s functions for the left(right) subsystem and $\bos{\mathcal{T}}$ is a tunneling matrix,
\begin{equation}
\bos{\mathcal{T}}_{\bos{k},\bos{p}}=
\left( \begin{array}{cc}
\bo{T}_{\bos{k},\bos{p}}&0\\
0&-\bo{T}^*_{-\bos{k},-\bos{p}}
\end{array}\right)
\end{equation}
with
\begin{equation}
\bo{T}_{\bos{k},\bos{p}}=\left(
\begin{array}{cc}
T^{\uparrow \uparrow}_{\bos{k},\bos{p}}&T^{\uparrow \downarrow}_{\bos{k},\bos{p}}\\
T^{\downarrow \uparrow}_{\bos{k},\bos{p}}&T^{\downarrow \downarrow}_{\bos{k},\bos{p}}
\end{array}
\right).
\end{equation}

Furthermore, we will assume for simplicity that the magnetisation of
our system is weak, hence we assume that the magnetisation is
homogeneous.
 Introducing the basis
 $\bos{\phi}_{\bsk}^\dagger=\left[\xi^\dagger_{\bos{k},\uparrow},\xi^\dagger_{\bos{k},\downarrow},\xi_{-\bos{k},\uparrow},\xi_{-\bos{k},\downarrow}\right]$ 
and choosing the quantisation axis along the magnetisation vector at each side of the junction, we have the following effective action,
 \begin{equation}
 S_{\rm eff}=S^L+S^R+S^T
 \end{equation} 
 and the inverse Green`s functions are given by
 \begin{equation}
 \bos{\mathcal{G}}^{-1}_{L(R)}=
 \left(\begin{array}{cc}
 \bo{g}^{-1}_{L(R)}&-\bos{\Delta}_{L(R)}\\
 -\bos{\Delta}^\dagger_{L(R)}&\bo{g}^{*-1}_{L(R)}
 \end{array}\right),
 \end{equation}
 with $\bo{g}^{-1}_{L(R),\alpha\beta}=\left[-\im \omega_n+\vep_{\bos{k}}-\alpha M^{L(R)}\right]\delta_{\alpha,\beta}$.
 The direction  of the magnetisation does not enter in the Green«s functions for the isolated systems on the left and right, the angle between the magnetisations, and hence the quantisation axes, only enters through the tunneling elements since we have applied different rotations on the left and right. This difference shows in the part of the Hamiltonian where the left and right systems couple.
 
We now want to calculate the Ginzburg-Landau free energy for 
the junction, $F_J$. To this end, we expand \eqref{eq:freejunk} 
to the lowest order in the tunneling elements, 
 \begin{align}
 F_J\approx & 
\frac{1}{\beta} \Tr\left[\bos{\mathcal{G}}_L\bos{\mathcal{T}}\bos{\mathcal{G}}_R\bos{\mathcal{T}}\right].
 \end{align}
Next, we use the Dyson equation to find an expansion of $\bos{\mathcal{G}}$ i.e. $\bos{\mathcal{G}}\approx \bos{\mathcal{G}}_0+\bos{\mathcal{G}}_0\bos{\Sigma}\bos{\mathcal{G}}_0+...$ were
 \begin{equation}
 \bos{\mathcal{G}}_0=\left(
 \begin{array}{cc}
 \bo{g}&0\\
 0&-\bo{g}^*
 \end{array}
 \right)
 \end{equation}
 and
 \begin{equation}
 \bos{\Sigma}=
 \left(
 \begin{array}{cc}
 0&\bos{\Delta}\\
 \bos{\Delta}^\dagger&0
 \end{array}
 \right).
 \end{equation}
 Since we are assuming that the magnetisation is homogeneous, $\bos{\mathcal{G}}_0$ is local in $\bos{k}-$space, $\bos{\Sigma}$ however is nonlocal as we do not want to impose any restriction on the superconducting pairing state.
 
From the free energy of the junction we can find the current of various quantities associated with tunneling. To show this we introduce a Hamiltonian $H\left(q,p\right)$ which is a function of the canonically conjugate variables $q,p$, and recall that the velocity operator is given by $\hat{\dot{q}}=\frac{\partial H}{\partial p}$. Consider now the quantity $\frac{\partial F}{\partial p}=-\frac{1}{\beta}\frac{\partial}{\partial p}\ln\left[
 \Tr \ex{-\beta H}\right]=\frac{1}{Z}\Tr\left[\frac{\partial H}{\partial p}\ex{-\beta H}\right]=\fv{\dot{q}}$.
 Hence we are easily able to to find the current over the 
 junction of for instance electrons.
 Since $\left[\phi,N\right]=\im $, the current is than given by $I_J=-e\fv{\dot{N}}=e\frac{\partial}{\partial \phi}F_J$ and the contribution to the spin current coming from tunneling across the barrier, is given by
 $\fv{\dot{S}_n}=-\mu_B\frac{\partial}{\partial \theta}F_J$, 
 where $\left[\theta,S_n\right]=\im$, and $\theta$ is the angle 
 in the plane perpendicular to the direction $\hat{n}$.
 \subsection{Ferromagnetic case}
 We start with the term corresponding to a junction between two ferromagnetic metals,
 \begin{align}
 F_J^{(0)}=&\frac{1}{\beta}\sum_{\newatop{\bos{k},\bos{p}}{\omega_n,\omega_\nu}}
 \tr\left(\bos{\mathcal{G}}_{0 \bos{k}}\bos{\mathcal{T}}_{\bos{k},\bos{p}}
 \bos{\mathcal{G}}_{0 \bos{p}}\bos{\mathcal{T}}^\dagger_{\bos{k},\bos{p}}\right)\nonumber \\
 =&\frac{1}{\beta}\sum_{\newatop{\bos{k},\bos{p}}{\omega_n,\omega_\nu}}\tr\left\{
 \bo{g}_{\bos{k}}\bo{T}_{\bos{k},\bos{p}}\bo{g}_{\bos{p}}\bo{T}^\dagger_{\bos{k},\bos{p}}\right.
 \nonumber \\
 &\qquad\left.
 +
 \bo{g}_{-\bos{k}}^*\bo{T}^*_{-\bos{k},-\bos{p}}\bo{g}^*_{-\bos{p}}\bo{T}^t_{-\bos{k},-\bos{p}}
 \right\}.
\end{align} 
We observe that upon letting $\bos{k}\rightarrow-\bos{k}$ and $\bos{p}\rightarrow-\bos{p}$ the second term above is the complex conjugate of the first one, hence the free energy is two times the real part of the first term,
\begin{align}
F_J^{(0)}=\frac{2}{\beta}\sum \Re& \left[
g_{\bos{k}}^+g_{\bos{p}}^+\leng{T^{++}_{\bos{k},\bos{p}}}^2
+ g_{\bos{k}}^-g_{\bos{p}}^-\leng{T^{--}_{\bos{k},\bos{p}}}^2\right.\nonumber \\
&\left.+g_{\bsk}^+g_{\bsp}^-\leng{T^{+-}_{\bsk,\bsp}}^2
+g_{\bsk}^-g_{\bsp}^+\leng{T^{-+}_{\bsk,\bsp}}^2
\right].
\end{align}
Here + and - means parallel or anti parallel to the magnetisation respectively. The angle between the magnetisations on the left and right enters through the tunneling elements $\leng{T^{\alpha\beta}_{\bsk,\bsp}}^2=1/2\left(1+\alpha\beta\hat{\bos{M}}^L\cdot\hat{\bos{M}}^R\right)\leng{T_{\bsk,\bsp}}^2$, when we assume that the spin in preserved across the tunnel barrier. The potential difference between the left and right subsystems are taken care of through the tunneling elements i.e. we assume that $\bom{T}_{\bsk,\bsp}$ is nonlocal in $\omega$-space with a boson frequency $\omega_\nu$, later we do an analytical continuation $\omega_\nu\rightarrow eV+\im\delta$. The structure of the Green`s functions is,
\begin{equation}
g_{\bsk}^\alpha g_{\bsp}^\beta=\frac{1}{-\im\left(\omega_n-\omega_\nu\right)+\vep_{\bsk \alpha}}
\frac{1}{-\im\omega_n +\vep_{\bsp \beta}},
\end{equation}
were $\vep_{\bsk\alpha}=\vep_{\bsk}-\alpha M^L$ and similarly for the right subsystem.
The zero`th order term of the free energy may be written in two parts, one part independent of the direction of the magnetisations and one part proportional to the dot product of the directions of the magnetisations, that is, to the cosine of the angle between the magnetisations, $F_J^{(0)}=\tilde{F}_J^{(0)}+\hat{\bos{M}}^L\cdot\hat{\bos{M}}^R\tilde{F}_{J,M}^{(0)}$.
With
\begin{align}
\tilde{F}_J^{(0)}=&\frac{1}{\beta}\sum_{\bsk,\bsp}\Re\left[
g_{\bsk}^+g_{\bsp}^+ +g_{\bsk}^-g_{\bsp}^-+g_{\bsk}^+g_{\bsp}^-+g_{\bsk}^-g_{\bsp}^+
\right]\leng{T_{\bsk,\bsp}}^2\nonumber \\
=&\leng{T}^2\int\de\xi_k\int\de\xi_pN(\xi_k)N(\xi_p)
\nonumber \\
&\left\{
\frac{f(\xi_k-M^L)-f(\xi_p-M^R)}{eV+\xi_k-\xi_p+M^R-M^L}\right.\nonumber\\
&+\frac{f(\xi_k-M^L)-f(\xi_p-M^R)}{eV+\xi_k-\xi_p+M^L-M^R}\nonumber \\
&+\frac{f(\xi_k-M^L)-f(\xi_p+M^R)}{eV+\xi_k-\xi_p-M^L-M^R}\nonumber \\
&+\left.\frac{f(\xi_k+M^L)-f(\xi_p-M^R)}{eV+\xi_k-\xi_p+M^L+M^R}
\right\}
\end{align}
and
\begin{align}
\tilde{F}_{J,M}^{(0)}=&\frac{1}{\beta}\sum\left[g_{\bsk}^+g_{\bsp}^+ + g_{\bsk}^-g_{\bsp}^- - g_{\bsk}^+g_{\bsp}^- - g_{\bsk}^-g_{\bsp}^+\right] \nonumber \\
=&\leng{T}^2\int\de\xi_k\int\de\xi_pN(\xi_k)N(\xi_p)\nonumber \\
&\left\{
\frac{f(\xi_k-M^L)-f(\xi_p-M^R)}{eV+\xi_k-\xi_p+M^R-M^L}\right.\nonumber \\
&+\frac{f(\xi_k-M^L)-f(\xi_p-M^R)}{eV+\xi_k-\xi_p+M^L-M^R}\nonumber \\
&-\frac{f(\xi_k-M^L)-f(\xi_p+M^R)}{eV+\xi_k-\xi_p-M^L-M^R}\nonumber \\
&-\left.\frac{f(\xi_k+M^L)-f(\xi_p-M^R)}{eV+\xi_k-\xi_p+M^L+M^R}
\right\}.
\end{align}
Here $N(\xi)$ is the density of states and $f(x)=1/(1+\ex{\beta x})$ is the Fermi-Dirac distribution function.
The spin current $\fv{\dot{S}_n}$, were $n$ denotes the direction perpendicular to the plane spanned by the magnetisation vectors, is now easily found by taking the derivative with respect to the angle between the magnetisations, i.e.
\begin{align}
\label{eq:spin0}
\fv{\dot{S}_n}_0=&-\mu_B\frac{\partial}{\partial \theta}F_J^{(0)}\nonumber \\
=&\mu_B\tilde{F}_{J,M}^{(0)}\sin\theta
\end{align}
This is precisely the same result as found previously in Ref. \onlinecite{nogueira2004} via a different route.
\subsection{Ferromagnetic superconducting case}
In this subsection, we calculate the free energy which is first order in the superconducting gap function on both the left and right side, i.e. we consider the ferromagnetic superconducting state. 
We want to find the Josephson current and also the two particle contribution to $\fv{\dot{S}_n}$. The free energy which can give rise to Josephson current is,
\begin{align}
\label{eq:frie2}
F_J^{(2)}= &\frac{1}{\beta}\Tr\left[\bom{G}_{0 L}\bos{\Sigma}_L\bom{G}_{0 L}\bom{T}\bom{G}_{0 R}\bos{\Sigma}_R\bom{G}_{0 R}\bom{T}\right]\nonumber \\
=&\frac{1}{\beta}\sum\tr
\bom{G}_{0\bsk}\bos{\Sigma}_{\bsk,\bsk^\prime}\bom{G}_{0\bsk^\prime}\bom{T}_{\bsk^\prime,\bsp}
\bom{G}_{0\bsp}\bos{\Sigma}_{\bsp,\bsp^\prime}\bom{G}_{0\bsp^\prime}\bom{T}^\dagger_{\bsk,\bsp^\prime}.
\end{align}
We now change variables,
$\newatop{\bsk}{\bsk^\prime}\rightarrow\newatop{\bsk+\bsq/2}{\bsk-\bsq/2}$
and
$\newatop{\bsp}{\bsp^\prime}\rightarrow\newatop{\bsp+\tilde{\bsq}/2}{\bsp-\tilde{\bsq}/2}$
were $\bsq(\tilde{\bsq})$ is the centre of mass momentum of the
Cooper-pairs on the left(right), and hence it is small.
 Expanding in $\bsq(\tilde{\bsq})$ amounts to an finding derivatives,
 we are not interested in derivatives and neglect the
 $\bsq(\tilde{\bsq})$ 
dependence in all terms except $\Sigma$, since we want to have the possibility $\Delta_{+-}\neq 0$. With these assumptions the free energy reads,
\begin{align}
F_J^{(2)}=&\frac{1}{\beta}\sum\tr\left[\bo{g}_{\bsk}\bos{\Delta}(\hat{\bsk},\bsq)\bo{g}^*_{-\bsk}
\bo{T}^*_{-\bsk,-\bsp}\bo{g}_{-\bsp}^*\bos{\Delta}^\dagger(\hat{\bsp},-\tilde{\bsq})\bo{T}^\dagger_{\bsk,\bsp}
\right.\nonumber \\
&\left.+ \bo{g}^*_{-\bsk}\bos{\Delta}(\hat{\bsk},-\bsq)\bo{g}_{\bsk}
\bo{T}_{\bsk,\bsp}\bo{g}_{\bsp}\bos{\Delta}(\hat{\bsp},\tilde{\bsq})\bo{g}^*_{\bsp}
\bo{T}^t_{-\bsk,-\bsp}\right].
\end{align}
Here we observe that upon changing the sign of all impulses in the
second term and using that 
$\bos{\Delta}(-\hat{\bsk})=-\bos{\Delta}(\hat{\bsk})$ for a triplet
superconductor, the second term is just the complex conjugate of the
first one.
  Hence, the free energy is just two times the real part of the first term. After some straightforward algebra, we obtain
\begin{widetext}
\begin{align}
\label{eq:josephfrie2}
F_J^{(2)}=-\leng{T}^2\sum_{\bsq,\tilde{\bsq}}
&\left\{ 
\left(1+\hat{\bos{M}}^L\cdot\hat{\bos{M}}^R\right)
\left[\left(A_{\uu}^{\uu}F_{\uu}^{\uu}+A_{\dd}^{\dd}F_{\dd}^{\dd}+A_{\ud}^{\ud}(F_{\ud}^{\ud}+F_{\du}^{\du})\right)\cos \Delta\phi \right.\right.\nonumber \\
&\qquad\qquad\qquad\qquad\left.+\left(A_{\uu}^{\uu}h_{\uu}^{\uu}+A_{\dd}^{\dd}h_{\dd}^{\dd}+A_{\ud}^{\ud}(h_{\ud}^{\ud}+h_{\du}^{\du})\right)\sin\Delta\phi\right]\nonumber\\
+&\left(1-\hat{\bos{M}}^L\cdot\hat{\bos{M}}^R\right)
\left[\left(
A_{\uu}^{\dd}F_{\uu}^{\dd}+A_{\dd}^{\uu}F_{\dd}^{\uu}-A_{\ud}^{\ud}(F_{\du}^{\ud}+F_{\ud}^{\du})
\right)\cos\Delta\phi\right.\nonumber \\
&\qquad\qquad\qquad\qquad\left.+\left(
A_{\uu}^{\dd}h_{\uu}^{\dd}+A_{\dd}^{\uu}h_{\dd}^{\uu}-A_{\ud}^{\ud}(h_{\du}^{\ud}+h_{\ud}^{\du})
\right)\sin\Delta\phi\right]\nonumber \\
+&\leng{\hat{\bos{M}}^L\times\hat{\bos{M}}^R}\left[\left(
A_{\uu}^{\ud}(F_{\uu}^{\ud}+F_{\uu}^{\du})-A_{\ud}^{\uu}(F_{\du}^{\uu}+F_{\ud}^{\uu})
+A_{\ud}^{\dd}(F_{\ud}^{\dd}+F_{\du}^{\dd})-A_{\dd}^{\ud}(F_{\dd}^{\ud}+F_{\dd}^{\du})\right)\cos\Delta\phi\right.\nonumber\\
&\qquad\qquad\left.\left.+
\left(
A_{\uu}^{\ud}(h_{\uu}^{\ud}+h_{\uu}^{\du})-A_{\ud}^{\uu}(h_{\du}^{\uu}+h_{\ud}^{\uu})
+A_{\ud}^{\dd}(h_{\ud}^{\dd}+h_{\du}^{\dd})-A_{\dd}^{\ud}(h_{\dd}^{\ud}+h_{\dd}^{\du})
\right)
\sin\Delta\phi\right]
\right\}
\end{align}
were we have introduced  the notation 
\begin{equation}
\label{eq:josephamp}
A_{\alpha\beta}^{\lambda\gamma}(\bsq,\tilde{\bsq})   
\equiv
2\int_{\Omega_{\bsk}>0}\frac{\de\Omega_{\bsk}}{4\pi}\int_{\Omega_{\bsp}>0}\frac{\de\Omega_{\bsp}}{4\pi}\leng{\Delta_{\alpha\beta}^L}\leng{\Delta_{\lambda\gamma}^R}
\leng{T_{\Omega_{\bsk},\Omega_{\bsp}}}^2,
\end{equation}  
\begin{equation}
F_{\alpha\beta}^{\lambda\gamma} \equiv
\int\de\xi_{\bsk}\de\xi_{\bsp}
\frac{\left(N(-\xi_{\bsk}+\alpha M^L)+N(\xi_{\bsk}+\beta M^L)\right)
\left(N(\xi_{\bsp}+\lambda M^R)+N(-\xi_{\bsp}+\gamma M^R)\right)
}{
4\left(\xi_{\bsk}-\frac{\alpha-\beta}{2}M^LÚ\right)\left(\xi_{\bsp}+\frac{\lambda-\gamma}{2}M^R\right)
}
\frac{
f(\xi_{\bsk})-f(\xi_{\bsp})
}{
eV+\xi_{\bsk}-\xi_{\bsp}
},
\end{equation}
and
\begin{align}
h_{\alpha\beta}^{\lambda\gamma} \equiv
\frac{\pi}{4}\int\de\xi_{\bsk}\left[f(\xi_{\bsk})-f(\xi_{\bsk}+eV)\right]
&
\left\{
\frac{N(\xi_{\bsk}+\alpha M^L+eV)
\left(N(\xi_{\bsk}+\gamma M^R)+N(-\xi_{\bsk}+\lambda M^R)\right)
}{
\left(\xi_{\bsk}+\frac{\alpha-\beta}{2}M^L+eV\right)
\left(\xi_{\bsk}-\frac{\lambda-\gamma}{2}M^R\right)
}
\right. 
\nonumber \\ 
+&\left.
\frac{N(\xi_{\bsk}+\beta M^L)
\left(N(\xi_{\bsk}+\lambda M^R+eV)+N(-\xi_{\bsk}+\gamma M^R-eV)\right)
}{
\left(\xi_{\bsk}-\frac{\alpha-\beta}{2}M^L\right)
\left(\xi_{\bsk}+\frac{\lambda-\gamma}{2}M^R+eV\right)
}
\right\}.
\end{align}
In addition, we have assumed that
$\phi_{\alpha\beta}^L-\phi_{\lambda\gamma}^R=\Delta\phi$ is
independent of the spin indices, this can be done since any coupling
between the different components of the superconducting order
parameter will give a phase locking \cite{leggett1966}. Further since
$T_{\bsk,-\bsp}=0$ \cite{bruder1995} the sum over $\bsk$ and $\bsp$ i
Equation \eqref{eq:frie2} may
be restricted to positive values only, and hence there are no problems
involved in splitting the superconducting orderparameter into an
amplitude and a phase in Equation \eqref{eq:josephfrie2}.
The Josephson current is now found by simply taking the derivative of the free energy with respect to the phase difference $\Delta\phi$. This leads to the floowing expression for the
Josephson current
\begin{align}
I_{J} = e \leng{T}^2\sum_{\bsq,\tilde{\bsq}}
&\left\{ 
\left(1+\hat{\bos{M}}^L\cdot\hat{\bos{M}}^R\right)
\left[\left(A_{\uu}^{\uu}F_{\uu}^{\uu}+A_{\dd}^{\dd}F_{\dd}^{\dd}+A_{\ud}^{\ud}(F_{\ud}^{\ud}+F_{\du}^{\du})\right)\sin \Delta\phi \right.\right.\nonumber \\
&\qquad\qquad\qquad\qquad\left.-\left(A_{\uu}^{\uu}h_{\uu}^{\uu}+A_{\dd}^{\dd}h_{\dd}^{\dd}+A_{\ud}^{\ud}(h_{\ud}^{\ud}+h_{\du}^{\du})\right)\cos\Delta\phi\right]\nonumber\\
+&\left(1-\hat{\bos{M}}^L\cdot\hat{\bos{M}}^R\right)
\left[\left(
A_{\uu}^{\dd}F_{\uu}^{\dd}+A_{\dd}^{\uu}F_{\dd}^{\uu}-A_{\ud}^{\ud}(F_{\du}^{\ud}+F_{\ud}^{\du})
\right)\sin\Delta\phi\right.\nonumber \\
&\qquad\qquad\qquad\qquad\left.-\left(
A_{\uu}^{\dd}h_{\uu}^{\dd}+A_{\dd}^{\uu}h_{\dd}^{\uu}-A_{\ud}^{\ud}(h_{\du}^{\ud}+h_{\ud}^{\du})
\right)\cos\Delta\phi\right]\nonumber \\
+&\leng{\hat{\bos{M}}^L\times\hat{\bos{M}}^R}\left[\left(
A_{\uu}^{\ud}(F_{\uu}^{\ud}+F_{\uu}^{\du})-A_{\ud}^{\uu}(F_{\du}^{\uu}+F_{\ud}^{\uu})
+A_{\ud}^{\dd}(F_{\ud}^{\dd}+F_{\du}^{\dd})-A_{\dd}^{\ud}(F_{\dd}^{\ud}+F_{\dd}^{\du})\right)\sin\Delta\phi\right.\nonumber\\
&\qquad\qquad\left.\left.-
\left(
A_{\uu}^{\ud}(h_{\uu}^{\ud}+h_{\uu}^{\du})-A_{\ud}^{\uu}(h_{\du}^{\uu}+h_{\ud}^{\uu})
+A_{\ud}^{\dd}(h_{\ud}^{\dd}+h_{\du}^{\dd})-A_{\dd}^{\ud}(h_{\dd}^{\ud}+h_{\dd}^{\du})
\right)
\cos\Delta\phi\right]
\right\}.
\end{align}
Similarly, we find the following expression for the two-particle contribution to the spin current
\begin{align}
\label{eq:spin2}
\fv{\dot{S}_n}_2=-\leng{T}^2\sum_{\bsq,\tilde{\bsq}}
&\left\{\sin\theta\left[\left(A_{\uu}^{\uu}F_{\uu}^{\uu}+A_{\dd}^{\dd}F_{\dd}^{\dd}
+A_{\ud}^{\ud}(F_{\ud}^{\ud}+F_{\du}^{\du}+F_{\ud}^{\du}+F_{\ud}^{\du})
-A_{\uu}^{\dd}F_{\uu}^{\dd}-A_{\dd}^{\uu}F_{\dd}^{\uu}\right)\cos\Delta\phi
\right.\right.
\nonumber \\
\qquad&\left.+\left(A_{\uu}^{\uu}h_{\uu}^{\uu}+A_{\dd}^{\dd}h_{\dd}^{\dd}
+A_{\ud}^{\ud}(h_{\ud}^{\ud}+h_{\du}^{\du}+h_{\ud}^{\du}+h_{\ud}^{\du})
-A_{\uu}^{\dd}h_{\uu}^{\dd}-A_{\dd}^{\uu}h_{\dd}^{\uu}\right)\sin\Delta\phi\right]
\nonumber\\
&+\cos\theta\left[
\left(
A_{\ud}^{\uu}(F_{\du}^{\uu}+F_{\ud}^{\uu})
-A_{\uu}^{\ud}(F_{\uu}^{\ud}+F_{\uu}^{\du})
+A_{\dd}^{\ud}(F_{\dd}^{\ud}+F_{\dd}^{\du})
-A_{\ud}^{\dd}(F_{\ud}^{\dd}+F_{\du}^{\dd})
\right)\cos\Delta\phi\right.\nonumber\\
&+\left.\left.
\left(
A_{\ud}^{\uu}(h_{\du}^{\uu}+h_{\ud}^{\uu})
-A_{\uu}^{\ud}(h_{\uu}^{\ud}+h_{\uu}^{\du})
+A_{\dd}^{\ud}(h_{\dd}^{\ud}+h_{\dd}^{\du})
-A_{\ud}^{\dd}(h_{\ud}^{\dd}+h_{\du}^{\dd})
\right)\sin\Delta\phi
\right]\right\}.
\end{align}
\end{widetext}
The spin current is not well defined in the limit when $\theta\rightarrow 0$, since we are calculating the current along the direction perpendicular to the {\it plane} spanned by the two magnetisation directions.
We observe that the total spin current is the sum of Eq. \eqref{eq:spin0} and \eqref{eq:spin2}. Hence, Eq. \eqref{eq:spin2} is a contribution originating with the interplay between superconductivity and magnetism. It is seen to disappear when superconductivity is lost. Moreover, we observe that there is a term in equation \eqref{eq:spin2} that is proportional to $\cos \theta$. In Josephson currents there also exist a term that is proportional to $\cos\Delta\phi$, which however is dissipative and vanishes if the voltage across the junction is set to zero. In the case of the spin current, we find that the condition that must be fulfilled for the cosine term to go to zero, is given by $\leng{\Delta_{\alpha\beta}^L}=\leng{\Delta_{\alpha\beta}^R}$, $M^L=M^R$ and $eV=0$. Furthermore, it is seen that the cosine 
term is associated with flipping 
\emph{one} electron from (spin)state $\alpha$ on the left to $-\alpha$ on the right, hence the cosine term is proportional to $\leng{\Delta_{\alpha\alpha}^L}\leng{\Delta_{-\alpha\alpha}^R}$ and similarly with the superscripts interchanged.

\section{\label{sec:conclude} CONCLUDING REMARKS}
We have derived a Ginzburg-Landau functional from a microscopic
Hamiltonian consisting of three terms, free fermigas, an Heisenberg
term and a spin generalised BCS term. We find two order parameters,
local magnetisation and the superconducting gap. We expand to fourth
order in the order parameters. The lowest order
coupling between the two order parameters is a third order term, this
term is only non zero when the material is in a non unitary
superconducting state i.e. $\bos{d}^*\times\bos{d}\neq 0$. From
an exact calculation of the coefficient of the third order term we
find that coexistence of ferromagnetism and superconductivity is
enhanced if the gradient of the density of states, $\left.\frac{\de}{\de\xi}N(\xi)\right\vert_{\xi=0}$, is \emph{positive} at Fermi level. 

We have also computed the tunneling contribution to the Ginzburg-Landau free energy. From this, we have found 
expressions for spin- and charge currentws in the spin- 
and charge channels across a tunneling junction separating 
two ferromagnetic $p$-wave superconductors. 

Superconductivity coexisting with ferromagnetism can be 
triggered by the magnetisation as a result of the presence of a third order term
\cite{shopova2005}, i.e. superconductivity appears at a higher
temperature than if magnetisation were not present. The sign of the
third order term is given by the the sign of the gradient of the DOS
at Fermi level. Since the Fermi level may be tuned by applying pressure, it may be possible to change the sign of the third order term by applying pressure. 
 
\begin{acknowledgements}
This work was supported by the Research Council of Norway Grants
No. 158518/431, No. 158547/431 (NANOMAT). E.K.D thanks NTNU for 
a research fellowship.
\end{acknowledgements}


\end{document}